\documentclass[]{emulateapj}
\usepackage{apjfonts}

\shorttitle{Continuous Gravitational Wave Detection Methods}
\shortauthors{Ellis et. al.}

\usepackage{amsmath,amssymb}
\usepackage{graphicx}
\usepackage{subfigure}
\newcommand{\be}{\begin{equation}}
\newcommand{\ee}{\end{equation}}
\newcommand{\omhat}{\hat{\Omega}}
\newcommand{\phat}{\hat{p}}
\newcommand{\hplus}{h_+}
\newcommand{\hcross}{h_{\times}}

\newcommand{\lp}{\left(}
\newcommand{\rp}{\right)}

\begin{document}


\title{Practical Methods for Continuous Gravitational Wave Detection using Pulsar Timing Data}


\author{J.~A. Ellis\altaffilmark{1,2}, F.~A. Jenet\altaffilmark{3}, and M.~A. McLaughlin\altaffilmark{2,4,5}}

\altaffiltext{1}{Center for Gravitation and Cosmology, University of Wisconsin Milwaukee, Milwaukee WI, 53211}
\altaffiltext{2}{Physics Deptartment, West Virginia University, Morgantown WV, 26505.}
\altaffiltext{3}{Center for Gravitational Wave Astronomy, University of Texas, Brownsville TX 78520.}
\altaffiltext{4}{Adjunct at the National Radio Astronomy Observatory, Green Bank, WV 24944.}
\altaffiltext{5}{Alfred P. Sloan Research Fellow.}


\begin{abstract}
Gravitational Waves (GWs) are tiny ripples in the fabric of space-time predicted by Einstein's General Relativity. Pulsar timing arrays (PTAs) are well poised to detect low frequency ($10^{-9}$ -- $10^{-7}$ Hz) GWs in the near future. There has been a significant amount of research into the detection of a stochastic background of GWs from supermassive black hole binaries (SMBHBs). Recent work has shown that single continuous sources standing out above the background may be detectable by PTAs operating at a sensitivity sufficient to detect the stochastic background. The most likely sources of continuous GWs in the pulsar timing frequency band are extremely massive and/or nearby SMBHBs. In this paper we present detection strategies including various forms of matched filtering and power spectral summing. We determine the efficacy and computational cost of such strategies. It is shown that it is  computationally infeasible  to use an optimal matched filter including the poorly constrained pulsar distances with a grid based method. We show that an Earth-term-matched filter constructed using only the correlated signal terms is both computationally viable and highly sensitive to GW signals. This technique is only a factor of two less sensitive than the computationally unrealizable optimal matched filter and a factor of two more sensitive than a power spectral summing technique. We further show that a pairwise matched filter, taking the pulsar distances into account is comparable to the optimal matched filter for the single template case and comparable to the Earth-term-matched filter for many search templates. Finally, using simulated data optimal quality, we place a theoretical minimum detectable strain amplitude of $h>2\times 10^{-15}$ from continuous GWs at frequencies  on the order $\sim1/T_{\rm obs}$.
\end{abstract}

\section{Introduction}
Low frequency ($10^{-9}$--$10^{-7}$ Hz) GWs are expected from supermassive black hole binary systems (SMBHBs), cosmic strings, the big bang and inflationary era of the early universe. GWs from these sources can manifest themselves in different ways. Single nearby SMBHBs can produce  resolvable waves  with periods on the order of years \citep{wl03,svv09,sv10}. SMBHBs and cosmic strings can also produce GW bursts \citep{dv01,smc07,lss09} in which the duration of the GW signal is much less than the observation time. We also expect PTAs to be sensitive to a stochastic background of unresolvable sources. Pulsar timing arrays (PTAs) offer an opportunity to detect low frequency GWs from all of these sources. The concept of a PTA composed of the best timed MSPs was first developed over two decades ago \citep{r89,fb90}. Today there are three main PTAs in existence with the goal of GW detection using pulsars: the European Pulsar Timing Array (EPTA; \citealt{jsk+08}), the North American Nanohertz Observatory for Gravitational waves (NANOGrav; \citealt{jfl+09}), and the Parkes Pulsar Timing Array (PPTA; \citealt{m08}), all of which are in collaboration to form the International Pulsar Timing Array (IPTA; \citealt{haa+10}). There is also a large international effort for the construction of future generation radio telescope arrays such as the Square Kilometer Array (SKA; \citealt{l09}), for which a primary science goal will be GW astrophysics.

Previous authors have developed statistical data-analysis methods for searches of the PTA data sets for stochastic backgrounds \citep{jhl+05,abc+08,hlm+09,ych+11,dfg+12} and burst sources \citep{fl10}. However, studies into continuous GW detection have been more theoretical or ``proof-of-principle'' in nature, as opposed to a more rigorous detection method aimed at real data-analysis pipeline implementation. Prior to the establishment of PTAs, \citet{jll+04} used existing pulsar data to rule out the proposed SMBHB system 3C66B, a possible source of continuous GWs. This work looked for the signature of a continuous GW in real pulsar data through the use of Lomb-Scargle periodograms and suggested a method for directed searches of known sources. \citet{yhj+10} also relied on the Lomb-Scargle periodogram to determine the sensitivity of a PTA to continuous GW sources as a function of GW frequency.  \citet{sv10} developed a Bayesian framework for the detection of continuous GWs from monochromatic SMBHBs in circular orbits. This work only included the earth term in the GW signal model and estimated the uncertainties one would expect on search parameters via the Fisher Information matrix, which is known to perform well in the high signal-to-noise ratio (SNR) regime \citep{v08}.  \citet{cc10} have developed a Bayesian Markov Chain Monte-Carlo (MCMC) technique for parameter estimation of an evolving SMBHB system in which the pulsar term is taken into account in the detection scheme. This work took advantage of a signal model in which the GW frequency evolves significantly in time to determine the pulsar distance. Most recently, \citet{lwk+11} have developed parameter estimation techniques based on vector Ziv-Zakai bounds incorporating the pulsar term and have placed limits on the detectable amplitude of a continuous GW. In their work, a future PTA with SKA sensitivity is assumed, resulting in  high SNR signals for their parameter estimation studies. All of these methods are promising for parameter estimation in a relatively high SNR (e.g. SNR=20 in \citet{cc10}) limit and a favorable signal model. 

The aim of this paper it to determine the most sensitive \textit{practical} detection technique for continuous GW sources in the PTA data making no assumptions about the SNR of the signal. This is done by comparing multiple detection techniques, using the minimum detectable amplitude as our figure of merit. The paper is ordered as follows. In section \ref{sec:method} we introduce the formalism and notation that we will use in the paper. In section \ref{sec:gw} we define the GW signal from a SMBHB and derive the resulting GW induced pulsar timing residuals. Section \ref{sec:mf} reviews methods of matched filtering, maximum likelihood detection techniques and power spectral summing. In section \ref{sec:data} we describe the simulated data sets that are used in this work. Section \ref{sec:analysis} describes the different detection techniques and discusses the main results of the paper. Finally, in section \ref{sec:summary} we summarize our work and mention prospects for future work.

\section{Method}
\label{sec:method}
\subsection{PTA Response to a Continuous GW}
\label{sec:gw}
While PTAs are poised to detect a stochastic GW background due to SMBHBs in the next five years, single, resolvable sources may also be detected at expected five-year sensitivity limits \citep{svv09}. A GW is defined as a metric perturbation to flat space time,
\be
h_{ab}(t,\omhat)=e_{ab}^+(\omhat)\hplus(t,\omhat)+e_{ab}^{\times}(\omhat)\hcross(t,\omhat),
\ee
where $\omhat$ is the unit vector pointing from the GW source to the Solar System Barycenter (SSB) and $\hplus$, $\hcross$ and $e_{ab}^A$ ($A=+, \times$) are the  polarization amplitudes and polarization tensors, respectively (See the Appendix for more details). The GW will cause a fractional shift in frequency, $\nu$, that can be defined by a redshift  in the times-of-arrival (TOAs)
\be
\frac{\delta\nu(t,\omhat)}{\nu}=- e^{A}_{ab}(\omhat)\frac{1}{2}\frac{\phat^a \phat^b}{1+\omhat\cdot\phat}\Delta h_A(t,\omhat),
\ee
where 
\be
\Delta h_A(t,\omhat)=h_A(t_e)-h_A(t_p).
\ee
Note that we use the standard Einstein summing convention. Here $t_e$ and $t_p$ are the time at which the GW passes the earth and pulsar respectively  and $\phat$ is the unit vector pointing from the SSB to the pulsar.  Henceforth, we will drop the subscript ``$e$'' denoting the earth time unless otherwise noted. From geometry we can write\footnote{Note that we use units in which $c=G=1$.}
\be
t_p=t-D(1-\cos\mu),
\ee
where $\omhat\cdot\phat=-\cos\mu$ and $D$ is the distance to the pulsar. To simplify our notation we introduce the pulsar ``antenna pattern functions''
\be
F^{A}(\omhat)=-\frac{1}{2}\frac{\phat^a \phat^b}{1+\omhat\cdot\phat}e^{A}_{ab}(\omhat),
\ee
where the redshift is now written as 
\be
\label{eq:redshift}
\frac{\delta\nu(t,\omhat)}{\nu}=F^{A}(\omhat)\Delta h_A(t,\omhat).
\ee
For this work we will only consider circular, non-precessing, monochromatic SMBHB systems, as these are expected to be the most prominent \citep{svv09}. Astrophysical justification for the above approximations can be found in \citet{sv10}. The word monochromatic indicates that the frequency evolution of the system is slow enough that we can make the approximation that $f(t_{p})=f(t_{e})={\rm const}$. It should be noted that we use the observed redshifted values. For example, the chirp mass and frequency in the rest frame are $\mathcal{M}_{r}=\mathcal{M}/(1+z)$ and $f_{r}=f_{0}(1+z)$, respectively, where $z$ is the cosmological redshift.  Assuming a monochromatic system with a circular orbit and an orbital angular frequency of $\omega_{\rm orb}=2(2\pi f_{0})=4\pi f_{0}$, where $f_{0}$ is the GW frequency we can now write the polarization amplitudes as
\begin{align}
\begin{split}
\hplus(t)&=h\big[(1+\cos^{2}\iota)\cos2\phi_{n}\cos(2\pi f_{0}t-\phi_{0}) \\
&-2\cos\iota\sin2\phi_{n}\sin(2\pi f_{0}t-\phi_{0})\big]
\end{split}\\
\begin{split}
\hcross(t)&=h\big[(1+\cos^{2}\iota)\sin2\phi_{n}\cos(2\pi f_{0}t-\phi_{0})\\
&+2\cos\iota\cos2\phi_{n}\sin(2\pi f_{0}t-\phi_{0})\big],
\end{split}
\end{align}
where $\phi_{0}$ is the orbital phase of the binary at $t=0$, $\iota$ is the inclination angle, and $\phi_{n}$ is the angle to the line of nodes. We define the amplitude $h$,
\be
\label{eq:amplitude}
h=2\frac{\mathcal{M}^{5/3} (\pi f_{0})^{2/3}}{D_{c}}.
\ee
The magnitude of the polarization amplitudes are proportional to the SMBHB chirp mass $\mathcal{M}=(M_{1}M_{2})^{3/5}/(M_{1}+M_{2})^{1/5}$, the comoving distance to the source $D_{c}$, and the GW frequency $f_{0}$ (twice the orbital frequency), and can be written as
\be
\begin{split}
\label{eq:apstrain}
h&\sim 8\times 10^{-15} \lp\frac{\mathcal{M}}{10^{9}\,M_{\odot}}\rp^{5/3}\lp\frac{D_{c}}{100 {\rm \,Mpc}}\rp^{-1}\\
&\times \lp\frac{f_{0}}{5\times 10^{-8}{\rm \,Hz}}\rp^{2/3}.
\end{split}
\ee
We now use the redshift given in Eq. \ref{eq:redshift} to compute the GW induced pulsar timing residuals
\be
\label{eq:res}
\begin{split}
r(t,\omhat)&=\int_{0}^{t} \frac{\delta\nu(t,\omhat)}{\nu}dt=\frac{1}{2\pi f_{0}}F^{A}(\omhat)\Delta h_A(t,\omhat)\\
&=\frac{1}{2\pi f_{0}}F^{A}(\omhat)\left[h_{A}(t_{e})-h_{A}(t_{e}-D(1+\omhat\cdot\phat))\right].
\end{split}
\ee
Here we have written out the explicit dependence on the pulsar distance and sky location. The first term in square brackets is the so-called  ``earth term'' because it refers to the metric perturbation at earth and is correlated in all sets of pulsar timing residuals. The second term in the square brackets is the so-called  ``pulsar term'' because it  refers to the metric perturbation at the pulsar and is uncorrelated in all sets of pulsar timing residuals. Notice that the pulsar term carries a dependence on the GW sky location $\omhat$, and \emph{all} of the dependence on the pulsar distance, $D$. As with the amplitude, we can express the approximate amplitude of the GW induced timing residuals as a function of the SMBHB source parameters \citep{svv09},
\be
\begin{split}
r&\sim25.7\,{\rm ns}\,\lp\frac{\mathcal{M}}{10^{9}\,M_{\odot}}\rp^{5/3}\lp\frac{D_{c}}{100 {\rm \,Mpc}}\rp^{-1}\\
&\times \lp\frac{f_{0}}{5\times 10^{-8}{\rm \,Hz}}\rp^{-1/3}.
\end{split}
\ee
 The residuals can be written as a function of $8+M$ parameters
\be
\label{eq:lambda}
\vec{\lambda}=\{\theta,\phi,\mathcal{M},D_{c},f_{0},\iota,\phi_{n},\phi_{0},\vec{D}\},
\ee
where $\vec{D}$ is a vector of the $M$ pulsar distances. Detecting and characterizing a signal that is a function of many parameters can be quite difficult and will be discussed in future papers. In this work we aim to give a baseline to the problem, in that, we will assume that all parameters are known and simply assess how well a particular detection method can confidently detect the signal.

\subsection{Matched Filtering and Power Spectral Summing}
\label{sec:mf}
Here we will outline matched filtering in terms of the log-likelihood, and a power spectral summing technique. First we will review matched filtering basics in the context of a PTA. The problem of detecting a signal in noisy data is well studied \citep{wz62}.   We assume that the noise in each pulsar is additive, stationary and gaussian. For this case, the data for each set of pulsar timing residuals $x_{\alpha}(t_{i})=x_{i\alpha}$ can be written as 
\be
x_{i\alpha}=r_{i\alpha}+n_{i\alpha},
\ee
where $r_{i\alpha}=r_{\alpha}(t_{i})$ and $n_{i\alpha}=n_{\alpha}(t_{i})$ are the signal and the noise in each data set. Here $i$ refers to the time index and $\alpha$ refers to the pulsar number. As is the method in matched filtering, we want to compare our data to a signal template of known form. Here we define $r_{i\alpha}=r_{\alpha}(t_{i},\vec\lambda)$ as our template of known form where $\vec\lambda$ is the vector of search parameters given in Eq. \ref{eq:lambda}.

We define the inner product of two functions of time $x(t_{i})$ and $y(t_{i})$ as
\be
(x|y)=\sum_{i,j}x_{i}(C)^{-1}_{ij}y_{j},
\ee
where $C$ is the covariance matrix of the noise. With this framework in place we can specialize to the case of our PTA data and templates with white noise. We find the inner product of our full set of residual data with the corresponding set of templates for that data set as
\be
\label{eq:wip}
(x|r(\vec\lambda))=\sum_{\alpha}\sum_{i}\frac{x_{i\alpha}r_{i\alpha}}{\sigma^{2}_{\alpha}},
\ee
where $\sigma^{2}_{\alpha}$ is the rms of the residuals from the $\alpha$th pulsar.  A Wiener optimal statistic can be defined as
\be
\label{eq:wiener}
\rho(\vec{\lambda})=\frac{(x|r(\vec{\lambda}))}{\sqrt{(r(\vec{\lambda})|r(\vec{\lambda}))}}.
\ee
If the noise is Gaussian and the signal is present, then the signal-to-noise ratio is given as
\be
\label{eq:snrw}
{\rm SNR}=\langle\rho(\vec{\lambda}^{\prime})\rangle=\sqrt{(r(\vec{\lambda}^{\prime})|r(\vec{\lambda}^{\prime}))},
\ee
where $\vec{\lambda}^{\prime}$ is the best estimate of the source parameters.  For our data analysis purposes we use the log-likelihood as our matched filtering statistic. Under the assumption of gaussian noise, we define the likelihood function as the probability of the data $x_{\alpha}(t_{i})$ given some set of model parameters $\vec{\lambda}$
\be
p(x|\vec{\lambda})=C_{\rm norm}{\rm \,exp}\left[-\frac{1}{2}\lp(x-r(\vec{\lambda}))|(x-r(\vec{\lambda}))\rp \right],
\ee 
where $C_{\rm norm}$ is a normalization constant. We now define the relative likelihood as $\Lambda(\vec{\lambda})=p(x|\vec{\lambda})/p(x|0)$, where $p(x|0)$ is the probability of  the data given the null hypothesis. From this we define the log-likelihood function
\be
\label{eq:like}
\ln\,\Lambda(\vec{\lambda})=\lp (x|r(\vec{\lambda}))-\frac{1}{2}(r(\vec{\lambda})|r(\vec{\lambda})) \rp.
\ee
By defining the log-likelihood in terms of the relative likelihood, we incorporate hypothesis testing (whether the GW signal is present or not) and parameter estimation into one statistic. We can define the SNR in terms of the log-likelihood as follows
\be
\label{eq:snrl}
\langle \ln\,\Lambda(\vec{\lambda}^{\prime}) \rangle=\frac{1}{2}\langle\rho(\vec\lambda^{\prime})\rangle^{2}=\frac{1}{2}{\rm SNR}^{2}.
\ee

To determine if a signal is present and to determine the source parameters $\vec{\lambda}$, one would need to search parameter space to find the maximum value of the likelihood. This can be done through grid based methods, Nested Sampling, or Markov Chain Monte-Carlo (MCMC). For this work we are only concerned with detection of a source. In this case, to claim a detection, our statistic (log-likelihood) must be greater than a threshold value determined by a specified false alarm value. 

While the log-likelihood has the ability to simultaneously carry out detection, parameter estimation and hypothesis testing, we now describe a method aimed at detection. In this method, we simply calculate the power spectrum of each set of pulsar timing residuals and then sum the power weighted by the variance of each data set and look for the maximum value over all frequency bins. We define our detection statistic as
\be
\label{eq:power}
\mathcal{P}=\max_{f}\sum_{\alpha=1}^{M}\frac{S_{\alpha}(f)}{\sigma_{\alpha}^{2}},
\ee
where $S_{\alpha}(f)$ and $\sigma_{\alpha}^{2}$ are the one sided power spectrum and the variance of the $\alpha$th pulsar data set, respectively. A detection is claimed when the value of $\mathcal{P}$ is greater than some threshold value $\mathcal{P}_{0}$  corresponding to a false alarm rate.

\section{Simulated PTA Data Sets}
\label{sec:data}
The simulated PTA data sets used for this work represent a best case scenario when it comes to \emph{data quality}. While the quality of the data is optimal, the properties of the PTA (i.e. distances, sky location, rms, etc.) are meant to represent a realizable case. The array consists of up to 100 pulsars uniformly distributed in both the azimuthal angle $\phi$ and in the cosine of the polar angle $\cos\theta$. The pulsar distances are also drawn from a uniform random distribution ranging from 0.5--3 kpc. It should be noted that although there are known millisecond pulsars with distances less than 0.5 kpc, the simulated pulsar distances here are meant to represent an average for a typical PTA. The pulsar timing residuals are evenly spaced over 10 years with 250 TOAs for each pulsar simulating roughly bi--monthly observations, and rms values drawn randomly from a uniform sample ranging from 100--300 ns. The noise is simulated to be white, gaussian, additive, and stationary. In real pulsar timing data, the residuals will be  unevenly sampled and the noise may have various red components. Fortunately, recent work suggests that most NANOGrav appear to be mostly white with little to no red noise contributions \citep{delphine,exd+12} In addition, the pulsar timing residuals will not be stationary, as a quadratic must be fit out of the data to account for the spin-down of the pulsar. Specifically, our definition of the inner product in Eq. \ref{eq:wip} no longer holds as we will need to include the covariance matrix of the data and incorporate a linear operator that takes into account this fitting.  These are issues that will need to be addressed in order to make a fully functional data analysis pipeline for continuous GW searches and will be addressed in future papers. However, here we will deal with the simple case to illustrate the efficacy of the studied search techniques on a data set of optimal quality.

\section{Analysis}
\label{sec:analysis}
In this section we compare four different detection techniques and determine their efficacy in terms of a minimum detectable amplitude as a function of the number of pulsars in the array. The four detection methods are the full matched filter, the earth--term only matched filter, the pairwise matched filter, and a simple power spectral summing technique. For all of these methods we will assume that we know the parameters of the source exactly and will only be interested in the lowest amplitude that each method can detect. Though unrealistic in most astrophysical scenarios, this gives us a simple baseline for comparison of the four methods.

\subsection{Detection Methods}
\label{sec:detection}
The full matched filter includes both the earth and pulsar terms from Eq. \ref{eq:res} and is thus a coherent search technique. This is the optimal detection statistic for a continuous wave buried in  gaussian noise. However, this method has the major drawback that it is  computationally expensive to carry out in practice as the pulsar distances must be added as search parameters. For example, if we have an array of 20 pulsars and want to search over just 100 trial distances for each pulsar, then we will need to use at least $10^{40}$ templates. If one does not use grid based methods, this number will drop significantly. However, even using more advanced methods like Nested Sampling or MCMC, including the pulsar distance as a search parameter is still computationally expensive. In the low SNR regime, where the likelihood surface is relatively flat and noisy, it may be impossible to get an accurate pulsar distance estimate with finite computational resources.

\begin{figure}[!t]
  \begin{center}
    \subfigure{\label{fig:fit}\includegraphics[width=0.49\textwidth]{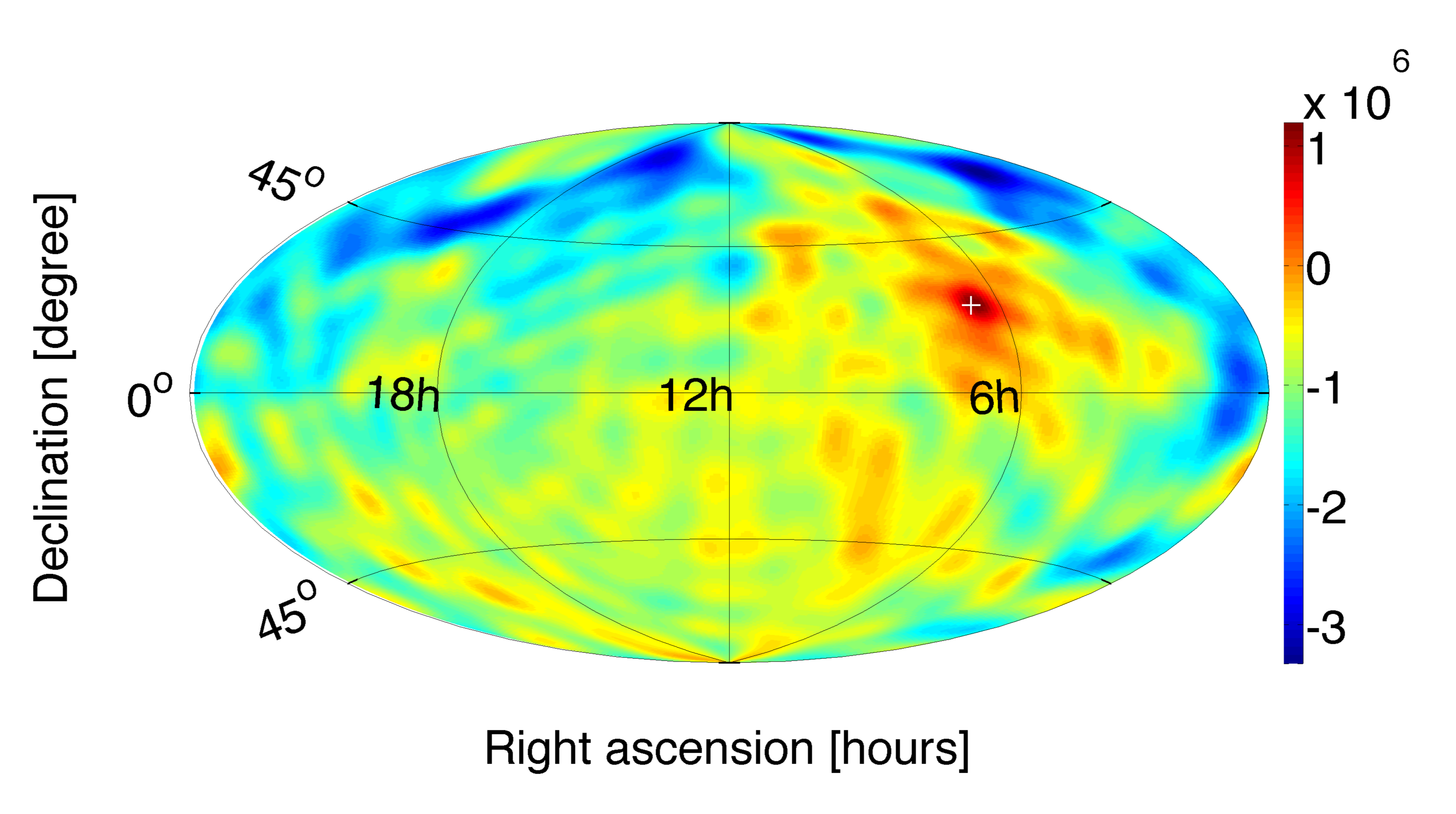}}
    \subfigure{\label{fig:gw}\includegraphics[width=0.49\textwidth]{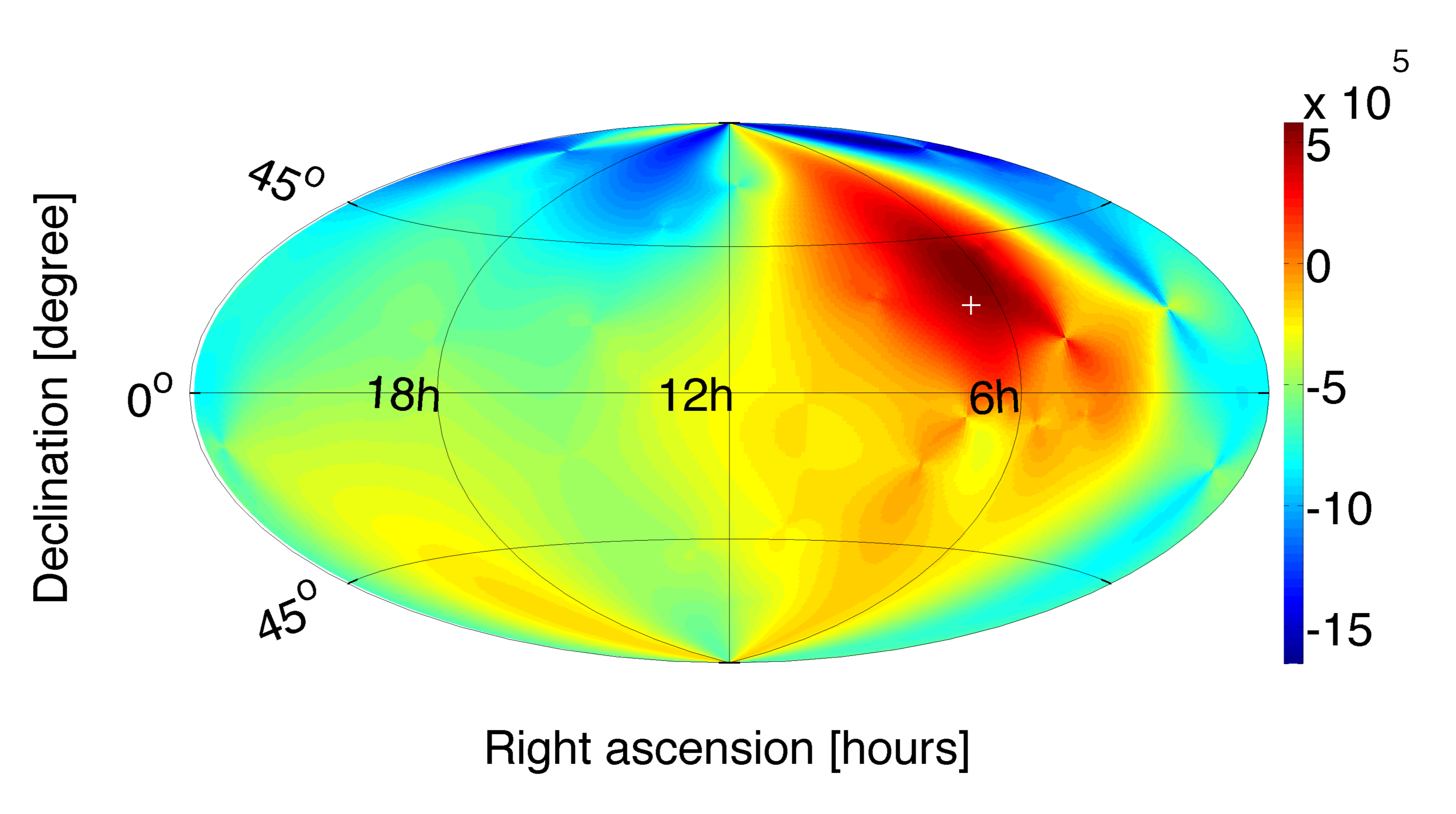}} \\
  \end{center}
  \caption{Skymaps created using the full matched filter (top) and the Earth-term-matched filter (bottom). The injected signal was very large (SNR=1000) for illustration purposes. The ``+'' symbol indicates the injected sky location. We can see that the sky localization is biased and has a large uncertainty for the Earth-term-matched filter as compared to the full matched filter. In this particular case we observe a 46\% loss in recovered SNR (See online version for color figures).}
    \label{fig:skymap}
\end{figure}

The Earth-term-matched filter uses templates that only depend on the coherent earth term and treats the pulsar term as a noise source.  The major advantage of using only the earth term from Eq. \ref{eq:res} for this detection method is that one does not need to include the pulsar distance in the search, making this method much less computationally expensive. Since we are not including the pulsar term in the analysis, we will always measure an SNR that is lower than the injected value. Also, there is a strong correlation between the pulsar distance and the sky location of the GW source. This will cause the recovered sky location to be biased and have larger uncertainties than the full matched filter case. This is illustrated in Figure \ref{fig:skymap} in  which a large GW (SNR=1000) is injected into the data from 40 pulsars. Both the full matched filter and the Earth-term-matched filter search over sky location, and a sky map is created where the color scale indicates the log-likelihood. We can see that the full matched filter does a good job of localization whereas the Earth-term-matched filter is biased with a much larger error box in the sky. For this realization, using the Earth-term-matched filter also results in a 46\% decrease in SNR compared to the full matched filter. 

The pairwise matched filter is a method that will allow us to take advantage of the pulsar term without the hindrance of an unworkable number of templates. This is done by constructing the full matched filter for each pair of pulsars, including the distance as a search term, and then adding the likelihoods as defined in Equation \ref{eq:like} in a pairwise fashion
\be
\ln\,\Lambda^{\rm PW}=\sum_{\alpha<\beta}(x|r(\vec\lambda))_{\alpha\beta}-\frac{1}{2}(x|r(\vec\lambda))_{\alpha\beta},
\ee
 where the $\alpha\beta$ subscript denotes an inner product using two pulsars and the sum indicates a sum over all $M(M-1)/2$ unique pulsar pairs. While we still need the same number of templates for the intrinsic parameters of the source, we need far fewer distance templates. For example, in the case of $M$ pulsars and 100 trial distances for each pulsar,  the full matched filter requires at least $10^{2M}$ distance templates. However, for the pairwise matched filter we only perform a coherent search each pulsar pair, therefore for each pair we only require $10^{4}$ distance templates making the total number of distance templates $M(M-1)/2\times 10^{4}$, which is \emph{significantly} less then the full matched filter. This method is still negatively affected by the strong degeneracy between pulsar distance and sky location but by using many pulsar pairs, this degeneracy will be greatly reduced.

The most simple and computationally inexpensive method that one could use to detect a continuous GW is power spectral summing as described in Eq. \ref{eq:power}. A method very similar to this was used in \citet{yhj+10} on real data to produce sensitivity curves for the PPTA pulsars. This method is relatively robust in that it does not depend on any signal model templates.   The disadvantages of this method are that it is incoherent because it does not keep track of phase information and it gives no indication of the true parameters of the source. With real data that is irregularly sampled and may contain red noise processes, going into the Fourier domain may pose problems that will be addressed in future work. However, for this analysis we are using this method as a baseline to compare the matched filtering statistics.

\begin{figure*}[!t]
  \begin{center}
  \includegraphics[width=1.0\textwidth]{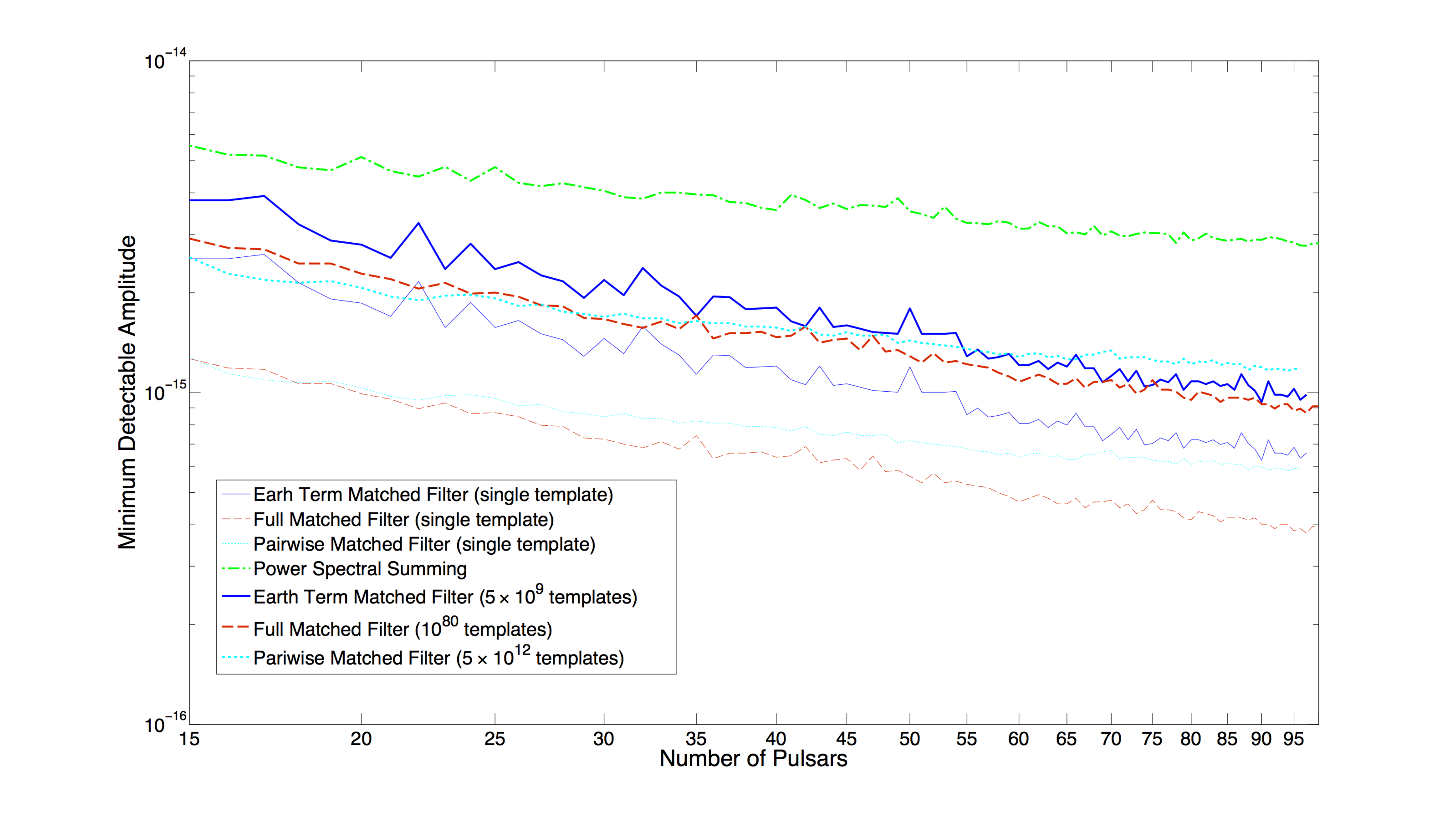}
   \end{center}
  \caption{Plot of minimum detectable amplitude vs. number of pulsars for the full matched filter (thin dashed line), Earth-term-matched filter (thin solid line), power spectral summing (thick dash-dot line) and the pairwise matched filter (thin dotted line). Also plotted are the full matched filter (thick dashed line) , the Earth-term-matched filter (thick solid line), and the pairwise matched filter (thick dotted line) with a realistic number of search templates. To make these plots, a Monte--Carlo simulation was run to find the amplitude at which 95\% of GW source realizations were detected for a given number of pulsars (See online version for color figures).}
    \label{fig:hmin}
\end{figure*}

\subsection{Efficacy of Detection Statistics}
\label{sec:results}
Here we will outline our Monte-Carlo simulations and present our results. A good figure of merit for a detection statistic is the minimum detectable amplitude, that is, we wish to find the amplitude that can be detected above some threshold in 95\% of the simulated realizations. To find the minimum detectable amplitude, we must first define a false alarm rate, that is, that rate at which we expect to make a detection when no signal is present. Assuming gaussian statistics, we want a $4\sigma$ detection significance corresponding to  a false alarm rate of 1/15,787. All four of our detection methods will have a different threshold value corresponding to this false alarm rate. To calculate these thresholds we perform the following simulation. We choose a template at random for each realization and calculate the detection statistic for 15,787 realizations of noise and record the maximum value. Statistically, this means that we would expect to get a value for our statistic that is above this maximum value $\sim$ 0.016\% of the time if our data was pure gaussian noise.  

A Monte-Carlo simulation was run to determine the minimum detectable amplitude as a function of the number of pulsars in the array for four detection statistics. The steps of the simulation are as follows: (\textit{i}) Choose the number of pulsars in the array and simulate residual data with white noise. (\textit{ii}) Fix\footnote{For this work we fix the frequency to the lowest detectable frequency of $f_{0}=1/T_{\rm obs}$ and the chirp mass to a reasonable value of $\mathcal{M}=5\times 10^{8}M_{\odot}$} the chirp mass $\mathcal{M}$, frequency $f_{0}$, and distance $D_{c}$ to construct the amplitude $h$ given in Equation \ref{eq:amplitude}. Then create a GW source in the sky with a given set of parameters $\vec{\lambda}=(\theta,\phi,\iota,\phi_{n})$ drawn from random distributions. (\textit{iii}) Add the GW induced residuals into the simulated PTA data. (\textit{iv}) Run the detection statistic code assuming all parameters are known exactly\footnote{This step is somewhat different for the Earth-term-matched filter because the largest SNR does not correspond to the case where the filter signal parameters are the same as the input signal parameters since we are not including the pulsar term. For this case we carry out a search over source sky location for each iteration in order to obtain the maximum possible log-likelihood.} and output the log-likelihood. If the log-likelihood is above a given threshold value, count as a detection, otherwise, count as a non-detection. (\textit{v}) draw a new parameter vector $\vec{\lambda}$ and repeat steps \textit{iii} and \textit{iv}. Repeat this for  10,000 GW source realizations (different realizations of $\vec{\lambda}$) and record the percent of sources detected. (\textit{vi}) Keep $\mathcal{M}$ and $f_{0}$ fixed and change $D_{c}$ to obtain a new amplitude $h$, repeat step \textit{v} until 95\% of the realizations are detected, (\textit{vii}) Change the number of pulsars in the array and repeat the entire procedure. In practice, a bisection root finding method is used to determine the detection probability instead of linearly increasing $h$ until the 95\% level is reached.

\subsubsection{Single Template Case}
We ran the simulation described above for each of our detection methods for PTAs with 15--100 pulsars. In this case we assume that there are no search templates except for the one exactly matching the data. This means that we will obtain the lowest possible false alarm probability. The results are shown in Figure \ref{fig:hmin}.  It is obvious that that the unrealizable full matched filter (thin dashed line) can detect the lowest amplitudes of the four detection methods tested. However, if one uses the pairwise matched filter (thin dotted line), very little sensitivity is lost for PTAs with up to 30 pulsars. Also, by using only the earth term in a matched filter search (thin solid line), the resulting minimum detectable amplitude is only a factor of two higher than the optimal method of filtering for the entire signal. It is also important to note that these two matched filtering methods (full and earth term) scale roughly the same with the number of pulsars in the array, therefore, this factor of two is independent of the number of pulsars. The incoherent power spectral summing method does approximately four times worse than the optimal matched filter case at 20 pulsars in the array and will continue to get worse as the number of pulsars in the array increases since the trend has a weaker dependence on the number of pulsars. It is known that the SNR scales as $\sqrt{M}$ for coherent methods and scales as $(M)^{1/4}$ for incoherent methods, where $M$ is the number of  ``detectors''.  The pairwise matched filter is also incoherent so the SNR has a flatter slope vs. N than when using the other two matched filtering methods. For a real PTA, the SNR will not scale exactly as mentioned above because each set of pulsar timing residuals do not have the exact same characteristics (different noise, sky location, distance, etc.). As a sanity check, we ran the simulation on PTAs where every pulsar had the same sky location, distance and rms residual. In this case the curves in Figure \ref{fig:hmin} scale exactly as mentioned above.

\subsubsection{Multiple Template Case}
The above section deals with the case of one template. In reality, the act of searching over many templates will serve to increase the false alarm probability. In the case of gaussian noise the false alarm rate can be calculated analytically as \citep{m07}
\be
\label{eq:pfa}
p_{\rm FA}=2\int_{\rho_{0}}^{\infty}d\rho \,e^{-\rho^{2}/2\sigma^{2}}=2\,{\rm erfc}(\rho_{0}/\sqrt{2}\sigma),
\ee
where $\rho_{0}$ is a threshold value of the SNR, $\rho$ is the output of the Wiener Filter in Eq. \ref{eq:wiener}, $\sigma$ is the standard deviation of the probability distribution function and ${\rm erfc}(z)$ is the complementary error function. Formally, this false alarm rate is derived based on a Wiener filter defined in Eq. \ref{eq:wiener}; however, since the expectation value of the log-likelihood is proportional the expectation value of the Wiener filter (see Eq. \ref{eq:snrl}), this equation is still valid for the log-likelihood method that we use here. Because we deal with only white gaussian noise in this paper, our simulated thresholds are in agreement with Eq. \ref{eq:pfa}. However when calculating a false alarm probability for a search with $N$ templates, the total false alarm probability is $P_{\rm FA}=Np_{\rm FA}$, when $p_{\rm FA}$ is much less than unity. This implies that if one wants to keep the same detection significance (less than 1/15,787 chance of occurring in noisy data), then the threshold value must be increased. When this is done, we can see from Figure \ref{fig:hmin} that the minimum detectable amplitudes for the Earth term matched filter, pairwise matched filter and full matched filter are nearly the same  because of the significantly smaller number of templates for the earth term and pairwise matched filters.  It is also important to note the values of $h$ on the $y$-axis. Since we are dealing with the best case scenario in terms of data quality (white gaussian noise, evenly spaced data, no timing model subtraction), this plot shows us that, at current levels of timing precision (rms $\sim$ 100 ns), we could never confidently detect any signal with an amplitude below $h=9\times 10^{-16}$ even with a PTA of 100 pulsars. At 20 pulsars, we could only possibly detect a source with $h>2\times10^{-15}$. In terms of placing limits on the minimum detectable amplitude for real data, our work suggests that if one implements an Earth-term-matched filtering technique as opposed to a power spectra summing technique as used in \citet{yhj+10}, one could improve the results by a factor of $\sim 2$.

\section{Summary}
\label{sec:summary}
In this work we have tested the efficacy of four detection techniques on simulated data sets when searching for continuous GW signals from SMBHBs. We have shown that a matched filter using only the correlated earth term in the search templates results in a minimum detectable amplitude that is 2 times higher than the optimal matched filter using both the correlated and uncorrelated terms. We have also shown that when using a pairwise matched filter it is possible, in principle, to obtain nearly the same sensitivity as the full matched filter. However, when performing a real search, the strong correlations between sky location of the GW source and pulsar distance may cause problems with SNR recovery, a problem that does not affect the Earth-term-matched filter. We have also shown that an incoherent power spectra summing method results in a minimum detectable amplitude that is 4 times higher at the present case of a 20 pulsar PTA and will continue to get worse as the number of pulsars in the array increases. When the number of search templates is taken into account, we find that using an Earth-term-matched filter is nearly as sensitive as the full matched filter.  Moreover, this work gives an idea of the prospects of detecting a continuous GW with PTAs, by placing lower limits on the detectable amplitude for data of optimal quality. The advantages and disadvantages of the various detection methods have been discussed and it has been shown that using a full matched filter with the pulsar distances included as search parameters is very computationally expensive (maybe even impossible for some cases). Because of this and the relatively low computational cost along with increased sensitivity over power spectrum techniques, the Earth-term-matched filter and pairwise matched filter appear to be practical choices for a detection method in a data analysis pipeline for use on real pulsar timing data. 

This work gives some insight into what detection techniques should be used in a fully functional pipeline. More sophisticated data analysis methods will have to include the effects of irregularly sampled data, red noise (both correlated noise in the form of the stochastic GW background and uncorrelated noise in the form of intrinsic timing noise and interstellar medium effects), and timing model parameter fits. The methods described here give basic detection algorithms that can be modified for use with real data. After a detection is made, the next step is parameter estimation. This will require fast, efficient algorithms to find the correct parameters in a large parameter space. Both of these challenges are currently being studied and will be the subject of future papers.

\appendix
\section{Polarization Tensor in the SSB Reference Frame}
Here we will show how one can convert the polarization tensors into the SSB coordinates.  Once again, a GW is defined as a metric perturbation to flat space time,
\be
h_{ab}(t,\omhat)=e_{ab}^+(\omhat)\hplus(t,\omhat)+e_{ab}^{\times}(\omhat)\hcross(t,\omhat),
\ee
where $\omhat$ is the unit vector pointing from the GW source to the Solar System Barycenter (SSB) and $\hplus$, $\hcross$ and $e_{ab}^A$ ($A=+, \times$) are the  polarization amplitudes and polarization tensors, respectively. The polarization tensors can be converted to the SSB by the following transformation. Following \citet{w87} we write the polarization tensors in terms of the wave principal axes described by unit vectors $\hat{m}$ and $\hat{n}$
\begin{align}
e_{ab}^+(\omhat)&=\hat{m}_a\hat{m}_b-\hat{n}_a\hat{n}_b,\\
e_{ab}^{\times} (\omhat)&=\hat{m}_a\hat{n}_b+\hat{n}_a\hat{m}_b.
\end{align}
In the SSB coordinate center we define
\begin{align}
\omhat &=-(\sin\theta\cos\phi)\hat{x}-(\sin\theta\sin\phi)\hat{y}-(\cos\theta)\hat{z},\\
\hat{m} &=-(\sin\phi)\hat{x}+(\cos\phi)\hat{y},\\
\hat{n} &=-(\cos\theta\cos\phi)\hat{x}-(\cos\theta\sin\phi)\hat{y}+(\sin\theta)\hat{z}.
\end{align}
In this coordinate system, $\theta=\pi/2-\delta$ and $\phi=\alpha$ are the polar and azimuthal angles of the source, respectively, where $\delta$ and $\alpha$ are declination and right ascension in usual celestial coordinates. In this coordinate system, the polarization tensors can be written as
\be
e^{+}(\theta,\phi)=
\begin{pmatrix}
\sin^{2}\phi-\cos^{2}\phi\cos^{2}\theta & -\cos^{2}\phi(1+\cos^{2}\theta)\sin\phi & \cos\phi\cos\theta\sin\theta \\
-\cos\phi(1+\cos^{2}\theta)\sin\phi & \cos^{2}\phi-\cos^{2}\theta\sin^{2}\phi & \cos\theta\sin\theta\sin\phi \\
\cos\phi\cos\theta\sin\theta & \cos\theta\sin\phi\sin\theta & -\sin^{2}\theta
\end{pmatrix}
\ee
\be
e^{\times}(\theta,\phi)=
\begin{pmatrix}
\cos\theta\sin2\phi & -\cos2\phi\cos\theta & -\sin\phi\sin\theta \\
-\cos2\phi\cos\theta & -2\cos\phi\cos\theta\sin\phi & \cos\phi\sin\theta \\
-\sin\phi\sin\theta & \cos\phi\sin\theta & 0
\end{pmatrix}.
\ee

\bibliographystyle{apj}
\bibliography{apjjabb,bib}

\end{document}